\documentclass[preprint,showpacs,preprintnumbers,amsmath,amssymb]{revtex4}

\usepackage{graphicx}
\usepackage{dcolumn}
\usepackage{bm}

\begin{document}


\title{Study of Coulomb interactions at VECC energies   
\\}

\author{Varinderjit Kaur}
\email{vkaur@thapar.edu}
\author{Suneel Kumar}%

\affiliation{%
School of Physics and Materials Science, Thapar University Patiala-147004, Punjab (India)\\
}
\author{Rajeev K. Puri}%
\affiliation{
Department of Physics, Panjab University, Chandigarh (India)\\
}%
\author{S. Bhattacharya}%
\affiliation{
Variable Energy Cyclotron Center, 1/AF, Bidhan Nagar, Kolkata INDIA\\
}%

\date{\today}
\maketitle
\baselineskip=18pt

\section{Introduction}
The possible nuclear reaction at SCC 500 developed in Variable energy Cyclotron Center (VECC) will
give us an insight to study the collective transverse flow which results due to two major interactions
(i) the attractive nuclear mean-field and (ii) the repulsive nucleon nucleon 
interactions \cite{aic}. Among various
observables and non observables in heavy-ion collisions, collective transverse flow
is one of the most sensitive and sought after phenomena at intermediate energies. The
collective transverse flow is the sideward deflection of the reaction products in 
phase space and is due to the interactions inside the reaction zone. 
This quantity has a beauty of vanishing at a certain incident energy. This energy is dubbed as balance 
energy (${E_{bal}}$) or the energy of vanishing flow (EVF) \cite{sood}. \\
Here, we have checked the influence of Coulomb interactions on the balance energy by taking
into account mass asymmetry. The mass asymmetry of the reaction can be
defined by the parameter ${{\eta}=  {\mid(A_T-A_P)}/{(A_T+A_P)\mid}}$; where ${A_T}$ and ${A_P}$ 
are the masses of
target and projectile respectively. The ${\eta}$ = 0 corresponds to the symmetric reactions, whereas, non-zero
value of ${\eta}$ define different mass asymmetries of the reaction. It is worth mentioning that the 
reaction dynamics in a symmetric reaction (${\eta}$ = 0) can be quite different compared to 
asymmetric reaction (${{\eta} \ne 0}$) \cite{vkaur}. 
The effect of the mass asymmetry of a reaction on the multifragmentation is studied many times in the literature
\cite{vkaur}. Unfortunately, very little study is available for the mass asymmetry of the reaction in
terms of transverse in-plane flow. Therefore, we study the effect of Coulomb interactions on 
balance energy for various colliding nuclei in terms of mass asymmetry. We plan to adress this question
using isospin-dependent quantum molecular dynamics (IQMD) model.\\
\section{The Model}
The model is the semi classical microscopic improved version of QMD model \cite{hartnack} which includes 
Skyrme forces, isospin-dependent Coulomb potential, Yukawa potential, symmetry potential, and NN 
cross-section. The details about the elastic and inelastic cross sections for 
proton-proton and neutron-neutron collisions can be found in Refs.\cite{hartnack}. Thus, the 
total interaction potential is given as: 
\begin{eqnarray}
V^{ij}(\vec{r}^\prime -\vec{r})&=&V^{ij}_{Skyrme}+V^{ij}_{Yukawa}+V^{ij}_{Coul}+\nonumber\\
& &V^{ij}_{mdi}+V^{ij}_{sym}
\end{eqnarray}\\
\section{Results and Discussion}
To check the effect of Coulomb interactions,  we have fixed 
(${A_{TOT}}$ = ${A_T +A_P}$ = 152) and varied the mass asymmetry of the reaction just like this: 
$_{26}Fe^{56}+_{44}Ru^{96}$ (${\eta = 0.2}$), $_{24}Cr^{50}+_{44}Ru^{102}$
(${\eta = 0.3}$), $_{20}Ca^{40}+_{50}Sn^{112}$ (${\eta = 0.4}$), $_{16}S^{32}+_{50}Sn^{120}$ 
(${\eta = 0.5}$), $_{14}Si^{28}+_{54}Xe^{124}$ (${\eta = 0.6}$), $_{8}O^{16}+_{54}Xe^{136}$ (${\eta = 0.7}$). 
Here, mass of the projectile varies between 16 and 56 units which is possible at present accelerator. We display in Fig.1,  mass asymmetry dependence of balance energy 
for hard and soft nuclear equation of state (NEOS) by switching off the Coulomb interactions. In addition, 
for a comparative study, the results in the presence of Coulomb interactions with soft NEOS are also shown. 
All the lines are fitted with power law of the form ${E_{bal} = C({\eta})^{\tau}}$, 
where C and ${\tau}$ are the constants. The values of ${\tau}$ in the absence of Coulomb 
interactions for soft and hard NEOS are 
0.375 and 0.282, respectively, while in the presence of Coulomb interactions for soft NEOS is ${\tau}$ = 0.06.   \\
If we compare the mass asymmetry dependence of balance energy with mass dependence, the trend is opposite
\cite{sood}. 
It is also clear from the figure, that shift in the balance energy is observed due to 
Coulomb interactions as well as due to different NEOS with mass asymmetry of the reaction. The shift
\begin{figure}
\includegraphics{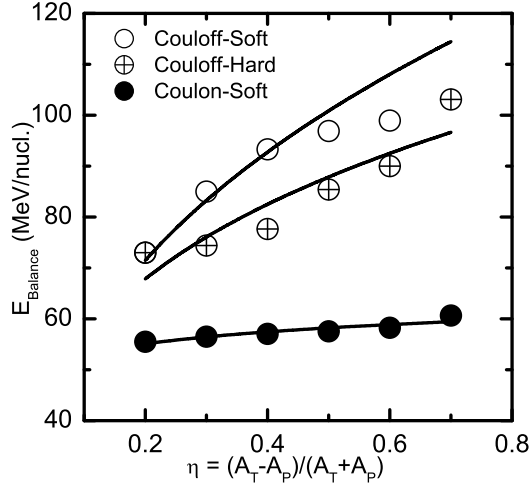}
\caption{\label{fig:1} Power law dependence of balance energy with mass asymmetry of the reaction using different equations of state.}  
\end{figure}
is more due to Coulomb interactions in comparison to NEOS, indicating the importance of 
Coulomb interactions in intermediate energy heavy-ion collisions. The higher balance energy is obtained
with Coulomb-off + soft NEOS followed by Coulomb-off + hard NEOS and 
finally Coulomb-on + soft NEOS. This study shows that the balance energy is affected by the Coulomb
interactions as well as different nuclear equations of state. The preliminary results calculated theoretically 
will be of great use for scientists at VECC. This study is further in progress.  

\end{document}